\documentclass[aps, amsmath, amssymb]{revtex4-2}

\usepackage{graphicx, color}
\usepackage{multirow, siunitx, url}

\DeclareMathOperator{\erf}{erf}

\begin{document}
\title{Analysis and application of multiplicative stochastic process with a sample-dependent lower bound}

\author{Ken Yamamoto}
\affiliation{Department of Physics and Earth Sciences, Faculty of Science, University of the Ryukyus, Senbaru, Nishihara, Okinawa 903-0213, Japan}
\author{Yoshihiro Yamazaki}
\affiliation{Department of Physics, Faculty of Advanced Science and Engineering, Waseda University, Okubo, Shinjuku, Tokyo 169-8555, Japan}

\begin{abstract}
A multiplicative stochastic process with the lower bound lognormally distributed is investigated.
For the process, the model is constructed, and its distribution function (involving four parameters) and the related statistical properties are derived.
By adjusting the parameters, it is confirmed that the theoretical distribution is consistent with empirical distributions of some real data.
\end{abstract}

\maketitle

\section{Introduction}
Probability distributions provide useful information on several complex phenomena and help achieve effective modeling.
Even if the precise dynamics or underlying mechanisms are unknown, a probability distribution can be simply constructed by observed data.
As such, the analysis of the empirical probability distributions is regarded as a possible first step toward understanding complex systems.

The heavy-tailed probability distribution, which decays slower than exponential distributions, is an important class of probability distributions~\cite{Foss}.
Power-law distribution~\cite{Newman} is representative of the heavy-tailed distributions.
The appearance of a heavy-tailed distribution implies that there exist elements having a far larger size than the characteristic size and that a mechanism for generating large-sized elements is embedded in the system.

Meanwhile, lognormal distribution is another representative heavy-tailed distribution that appears in various complex systems~\cite{Kobayashi, Limpert}.
Here, the random variable $X$ is said to follow a lognormal distribution if its logarithm $\ln X$ is normally distributed.
The probability density $f(x)$ and cumulative distribution $F(x)$ of the lognormal distribution can be described as follows:
\begin{subequations}\label{eq1}
\begin{align}
f(x)&=\frac{1}{\sqrt{2\pi}\sigma x}\exp\left(-\frac{(\ln x-\mu)^2}{2\sigma^2}\right),\label{eq1a}\\
F(x)&=\int_x^\infty f(x)dx = \frac{1}{2}\left[1-\erf\left(\frac{\ln x-\mu}{\sqrt{2}\sigma}\right)\right]\label{eq1b},
\end{align}
\end{subequations}
where $\mu$ and $\sigma$ correspond to the mean and standard deviation of $\ln X$, respectively, and $\erf$ is the Gauss error function defined by the following:
\[
\erf(x)=\frac{2}{\sqrt{\pi}}\int_0^x e^{-y^2} dy.
\]
The lognormal distribution is generated by the following multiplicative stochastic process
\begin{equation}
X_{n+1}=B_n X_n,
\label{eq2}
\end{equation}
where $B_0, B_1,\ldots$ represent positive random variables that are independently and identically distributed.
In fact, according to the central limit theorem, the distribution of $\ln X_n$ converges to a normal distribution.
Qualitatively, because the lognormal distribution is attained via numerous factors that are accumulated multiplicatively, the lognormal distribution is observed in a variety of situations.
This model was first introduced by Kolmogorov~\cite{Kolmogorov} to explain a fragment-size distribution in a cascade fracture.

The multiplicative stochastic process~\eqref{eq2} can produce a power-law distribution when an additional effect is introduced,
one example being the introduction of a lower bound $a$ for $X_n$~\cite{Levy} such that $X_n$ is restricted to not less than $a$:
\begin{equation}
X_{n+1}=\begin{cases}
B_n X_n & (B_nX_n\ge a)\\
X_n & (B_nX_n<a).
\end{cases}
\label{eq3}
\end{equation}
In the $n\to\infty$ limit, $X_n$ follows a stationary power-law distribution:
\begin{equation}
\Phi_a(x)=P(X_n>x)=\left(\frac{x}{a}\right)^{-\beta} \quad(x\ge a),
\label{eq4}
\end{equation}
where the exponent $\beta>0$ is given by the positive solution of
\[
E[B_t^\beta]=1.
\]
The corresponding probability density function is expressed as
\begin{equation}
\phi_a(x)=-\frac{d\Phi_a(x)}{dx}=\frac{\beta}{a}\left(\frac{x}{a}\right)^{-\beta-1} \quad(x\ge a).
\label{eq5}
\end{equation}
Historically, this model dates back to the work of Champernowne~\cite{Champernowne}, who applied it to the power-law income distribution.
A continuous version that involves geometric Brownian motion with a reflection boundary has been investigated~\cite{Gabaix}.
Other mechanisms for the power-law distributions from the multiplicative process~\eqref{eq2} have been also studied in terms of the introduction of additive noise~\cite{Takayasu}, a reset event~\cite{Manrubia}, random stopping~\cite{Yamamoto2012}, and the sum of variables~\cite{Yamamoto2014}.

In this paper, we investigate a stochastic process based on business-related empirical data.
Here, the empirical distribution of the data follows a lognormal distribution up to the mid-sized region, with a tail fatter than the usual lognormal distribution.
We propose that this tail behavior can be modeled via a multiplicative stochastic process with a specific lower bound, in which the initial value is drawn from a lognormal distribution and the lower bound is equal to this initial value.
After formulating the model, we derive the distribution function and its properties.

\section{Modeling}
\begin{figure}[t!]\centering
\includegraphics[scale=0.9]{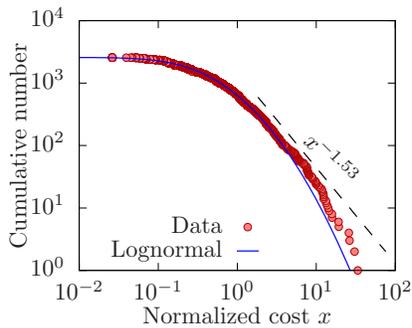}
\caption{
The cumulative number of troubleshooting costs (normalized by the average cost) in the company (the number of cases is $N=2571$).
The solid curve is the lognormal distribution with $\mu=-0.84$ and $\sigma=1.24$, and the dashed line indicates the power-law slope $x^{-1.53}$ in the tail.
}
\label{fig1}
\end{figure}

Figure~\ref{fig1} shows the cumulative number of troubleshooting costs in a specific Japanese company.
The cumulative number is defined as the number of cases with costs larger than $x$.
The absolute values are substituted with the costs divided by the average as shown in Fig.~\ref{fig1}~\cite{Company}.
The total number of cases is $N=2571$.
The solid curve corresponds to the lognormal distribution with $\mu=-0.84$ and $\sigma=1.24$.
(Note that the cumulative number is given by $NF(x)$.)
The empirical distribution deviates from the lognormal distribution in the tail ($x\gtrsim10$).

While only a few (the order of $10^1$) points deviate from the lognormal distribution, they are highly expensive cases, ranging from approximately $10$ to $100$ times greater than the average value.
Hence, the precise description of the tail behavior based on a mathematical mechanism will be financially important in practical situations.
Moreover, it has been widely observed that the main part of the data follows a lognormal distribution, but the tail deviates from the lognormal~\cite{Clementi, Golosovsky, Corral}.
Such tail behavior is generally difficult to determine and can often be controversial.
For example, while the population distribution of American cities was reported to be lognormally distributed~\cite{Eeckhout}, its tail was later noted to exhibit a power-law decay~\cite{Levy2009}.
This issue has not been resolved yet because the measurement of the tail is problematic~\cite{Eeckhout2009, Ioannides}.
Furthermore, the population distribution varies according to city, town, and village~\cite{Sasaki}, and a numerical model incorporating the natural and social effects on population increase or decrease in a simplified form has been proposed~\cite{Sasaki, Yamazaki}.
Therefore, the tail of an empirical distribution is worth modeling and analyzing in theoretical terms.

Assuming the tail of the empirical distribution shown in Fig.~\ref{fig1} follows power-law decay, we obtain the power-law exponent $-1.53\pm0.08$ using the method by Clauset, Shalizi, and Newman~\cite{Clauset}.
However, this power-law decay (shown by the dashed line in Fig.~\ref{fig1}) does not seem to express the tail of the empirical distribution well.
The power-law decay cannot always be justified theoretically, and it should be regarded as a rough assumption or a working hypothesis.
A more appropriate function may be required to understand the system accurately.

We thus propose the following phenomenological model.
Based on the lognormal distribution in the low- and mid-cost regions, it is assumed that the cost is determined by several multiplicatively affecting factors.
Thus, this can be naively modeled using the stochastic process described in Eq.~\eqref{eq2}, which produces the lognormal distribution.
Additionally, for expensive cases, additional tasks, such as individual inspection and customization, tend to be involved.
These additional tasks may increase the cost but never reduce it to less than the initial cost.
This operation can be expressed by a multiplicative process wherein the initial cost acts as the lower bound.
Due to this additional process, the price can further increase, resulting in a deviation from the lognormal distribution.

The above model can be simulated as follows.
First, let us define $x_*$ as the boundary cost where the empirical data begins to deviate from the lognormal distribution.
A random number $x$ is then drawn from the lognormal distribution.
If $x\ge x_*$, the following multiplicative process, wherein the lower bound is $x$ itself, is performed, where the initial value is $X_0=x$ and the evolution is given by
\[
X_{n+1}=
\begin{cases}
B_n X_n & (B_n X_n\ge x)\\
X_n & (B_n X_n < x)
\end{cases}
\]
until $X_n$ becomes stationary.
Here, $B_0, B_1,\ldots$ are independently and identically distributed positive random numbers.
Unlike in Eq.~\eqref{eq3}, where the lower bound $a$ is constant, the lower bound $x$ is distributed lognormally and varies for each sample in this model.

Therefore, $x$ is lognormally distributed for $x<x_*$, whereas the distribution changes for $x\ge x_*$ due to the additional process.
In the next section, we analyze the proposed model and derive the distribution function for $x\ge x_*$.

\section{Model analysis}\label{sec3}
As mentioned earlier, the proposed model involves a multiplicative process with a varying lower bound.
First, we derive the corresponding probability density $g(x)$ and cumulative distribution $G(x)$.

For $x<x_*$, the effect of the lower bound does not come into play.
Hence, $g(x)$ and $G(x)$ are identical to the lognormal distribution $f(x)$ and $F(x)$ described in Eq.~\eqref{eq1}, respectively.

\begin{figure}[t!]\centering
\includegraphics[clip]{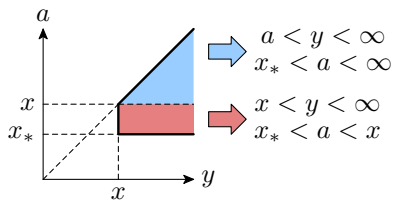}
\caption{
Illustration of the interchange of the integration in Eq.~\eqref{eq7}.
The shaded area composed of semi-infinite triangular (corresponding to $G_1(x)$) and rectangular ($G_2(x)$) areas represent the domain of the iterated integral in the $y$-$a$ plane.
}
\label{fig2}
\end{figure}

For $x\ge x_*$, $g(x)$ is obtained by superposing the stationary distributions corresponding to the different values of the lower bound.
When the lower bound is $a$, the stationary distribution follows the power law $\phi_a(x)$ described in Eq.~\eqref{eq5}.
In the proposed model, the lower bound $a$ has a lognormal weight $f(a)$ [see Eq.~\eqref{eq1a}], such that
\begin{equation}
g(x)=\int_{x_*}^x \phi_a(x)f(a) da.
\label{eq6}
\end{equation}
The cumulative distribution can be expressed as follows:
\[
G(x)=\int_x^\infty g(y)dy=\int_x^\infty dy\int_{x_*}^y \phi_a(y)f(a)da.
\]
The order of integration can be interchanged as follows (see Fig.~\ref{fig2} for reference):
\begin{equation}
\int_x^\infty dy\int_{x_*}^y da\phi_a(y)f(a)=\left(\int_x^\infty da\int_a^\infty dy+\int_{x_*}^x da\int_x^\infty dy\right)\phi_a(y)f(a)=G_1(x)+G_2(x).
\label{eq7}
\end{equation}
By the normalization $\int_a^\infty \phi_a(y)dy = 1$, we can easily obtain
\begin{equation}
G_1(x)=\int_x^\infty f(a) da\int_a^\infty \phi_a(y)dy = \int_x^\infty f(a) da = F(x).
\label{eq8}
\end{equation}
The function $G_2(x)$ can be written as
\begin{align}
G_2(x)&=\int_{x_*}^x da f(a) \int_x^\infty \phi_a(y)dy
=\int_{x_*}^x f(a)\Phi_a(x) da\nonumber\\
&=\int_{x_*}^x \frac{1}{\sqrt{2\pi}\sigma a}\exp\left(-\frac{(\ln a-\mu)^2}{2\sigma^2}\right)\left(\frac{x}{a}\right)^{-\beta} da\nonumber\\
&=x^{-\beta}\exp\left(\mu\beta+\frac{\sigma^2\beta^2}{2}\right)\int_{x_*}^x\frac{1}{\sqrt{2\pi}\sigma a}\exp\left(-\frac{(\ln a-\mu-\beta\sigma^2)^2}{2\sigma^2}\right)da.
\label{eq9}
\end{align}
Since the integrand represents the lognormal probability density (as a function of $a$) where the parameter $\mu$ is replaced with $\mu+\beta\sigma^2$, we have
\begin{equation}
G_2(x)=\frac{x^{-\beta}}{2}\exp\left(\mu\beta+\frac{\sigma^2\beta^2}{2}\right)\left[\erf\left(\frac{\ln x-\mu-\beta\sigma^2}{\sqrt{2}\sigma}\right)-\erf\left(\frac{\ln x_*-\mu-\beta\sigma^2}{\sqrt{2}\sigma}\right)\right].
\label{eq10}
\end{equation}
Thus, for $x\ge x_*$, we can obtain
\begin{equation}
G(x)=\frac{1}{2}\left[1-\erf\left(\frac{\ln x-\mu}{\sqrt{2}\sigma}\right)\right]
+\frac{1}{2}\exp\left(\mu\beta+\frac{\sigma^2\beta^2}{2}\right)x^{-\beta}
	\left[\erf\left(\frac{\ln x-\mu-\beta\sigma^2}{\sqrt{2}\sigma}\right)-\erf\left(\frac{\ln x_*-\mu-\beta\sigma^2}{\sqrt{2}\sigma}\right)\right].
\label{eq11}
\end{equation}
Here, the first term on the right-hand side represents the lognormal distribution, while the second term is regarded as the attendant correction term.

By differentiating Eq.~\eqref{eq11}, we obtain 
\begin{align*}
g(x)&=-\frac{dG(x)}{dx}\\
&=\frac{1}{2}\beta x^{-\beta-1}\exp\left(\mu\beta+\frac{\sigma^2\beta^2}{2}\right)
	\left[\erf\left(\frac{\ln x-\mu-\beta\sigma^2}{\sqrt{2}\sigma}\right)-\erf\left(\frac{\ln x_*-\mu-\beta\sigma^2}{\sqrt{2}\sigma}\right)\right]
\end{align*}
for $x\ge x_*$.

\begin{figure}[t!]\centering
\raisebox{40mm}{(a)}\includegraphics[scale=0.9]{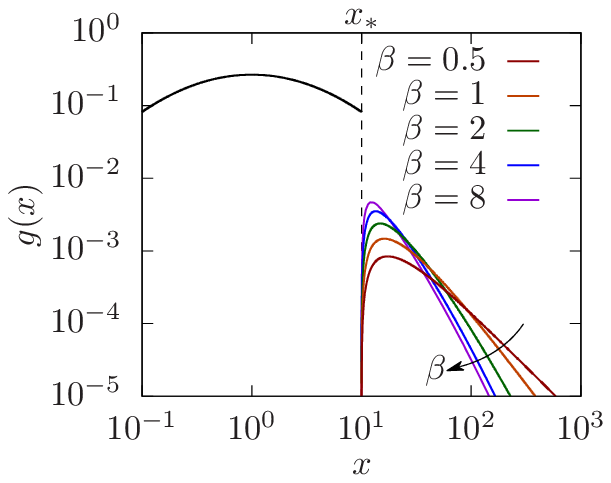}
\hspace{1cm}
\raisebox{40mm}{(b)}\includegraphics[scale=0.9]{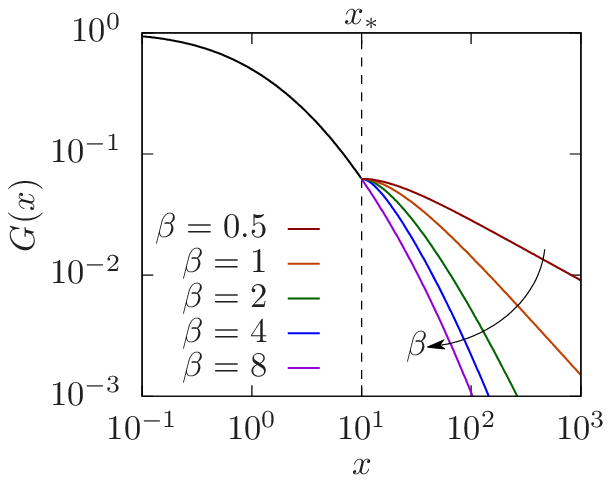}
\caption{
Graphs of (a) the probability density $g(x)$ and (b) the cumulative distribution $G(x)$ for $(\mu, \sigma)=(0, 1.5)$, $x_*=10$, and $\beta=0.5, 1, 2, 4$, and $8$.
The decay is fast for large $\beta$ (indicated by the arrow in each graph).
}
\label{fig3}
\end{figure}

Figure~\ref{fig3} shows graphs of $g(x)$ and $G(x)$ with fixed $(\mu,\sigma)=(0, 1.5)$ and $x_*=10$, and varying $\beta=0.5, 1, 2, 4$, and $8$.
The tail decays rapidly for large $\beta$, and the probability density $g(x)$ is not continuous at $x=x_*$: $g(x)\to0$ as $x\searrow x_*$.
In accordance with the discontinuity of $g(x)$, the $G(x)$ graph is not smooth at $x=x_*$.

If a random variable $X$ has the probability density $g(x)$, the $k$th moment of $X$ can be calculated as
\begin{align*}
E[X^k]=&\int_0^{x_*} x^k f(x)dx+\int_{x_*}^\infty x^k g(x) dx\\
&=\frac{1}{2}\exp\left(k\mu+\frac{k^2\sigma^2}{2}\right)\left[\frac{2\beta-k}{\beta-k}-\frac{k}{\beta-k}\erf\left(\frac{\ln x_*-\mu-k\sigma^2}{\sqrt{2}\sigma}\right)\right]
\end{align*}
for $k<\beta$, and the moment diverges for $k\ge\beta$.
The calculation of the integral is straightforward but cumbersome.
This criterion for the divergence of moments can be derived simply using the asymptotic power law $g(x)\sim x^{-\beta-1}$ or $G(x)\sim x^{-\beta}$.

\section{Data analysis}
The function $G(x)$ involves four adjustable parameters: $\mu, \sigma, x_*$, and $\beta$.
The optimal value of these parameters for fitting the given data can be determined as follows.
For fixed $x_*$, we can obtain $\mu$ and $\sigma$ according to the lognormal fitting in the region $x<x_*$.
Using these values $\mu$ and $\sigma$, we can obtain $\beta$ by fitting Eq.~\eqref{eq11} in the region $x>x_*$.
By calculating $\mu,\sigma,\beta$, and the fitting error for different $x_*$, we can determine the optimal set of the parameters as one minimizing the fitting error.

\begin{figure}[t!]\centering
\includegraphics[scale=0.9]{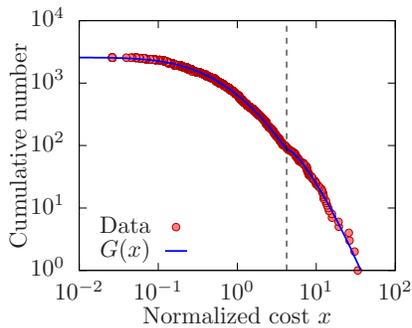}
\caption{
The cumulative number of normalized trouble handling costs (circle points) and $G(x)$ (solid curve).
The fitting parameters are $(\mu, \sigma)=(-0.84, 1.24)$, $x_*=4.21$, and $\beta=5.28$.
The vertical dashed line indicates the position of $x_*$.
}
\label{fig4}
\end{figure}

Figure~\ref{fig4} shows the result of the curve fitting of $G(x)$ to the empirical distribution shown in Fig.~\ref{fig1}.
Here, the parameter values are $(\mu, \sigma)=(-0.84, 1.24)$ (same as those in Fig.~\ref{fig1}), $x_*=4.21$, and $\beta=5.28$.
$G(x)$ is consistent with the empirical distribution almost over the entire range, which indicates the validity of the proposed model i.e., the multiplicative stochastic process with the sample-dependent lower bound.

The naive power-law fitting shown in Fig.~\ref{fig1} gives the power-law exponent $-1.53(>-2)$ for a cumulative distribution, implying that the variance diverges if the same power law extends to infinity.
In contrast, we obtained $\beta=5.28$ when using $G(x)$, which results in finite variance.
Here, it can be expected to obtain the correct statistical properties of the data when using $G(x)$, ensuring appropriate and effective cost management in practical situations.

\begin{figure}[t!]\centering
\raisebox{40mm}{(a)}\includegraphics[scale=0.9]{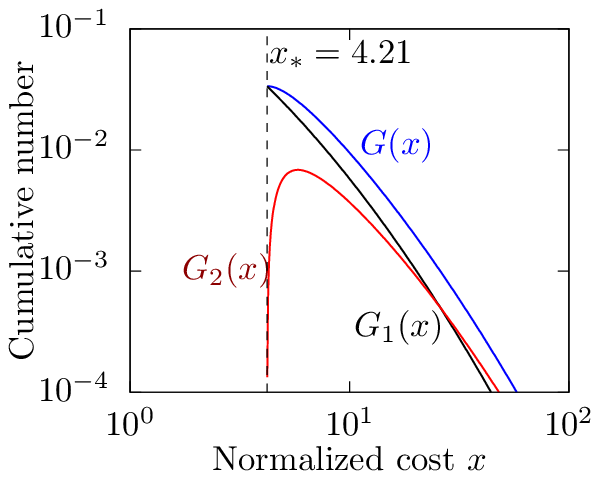}
\hspace{1cm}
\raisebox{40mm}{(b)}\includegraphics[scale=0.9]{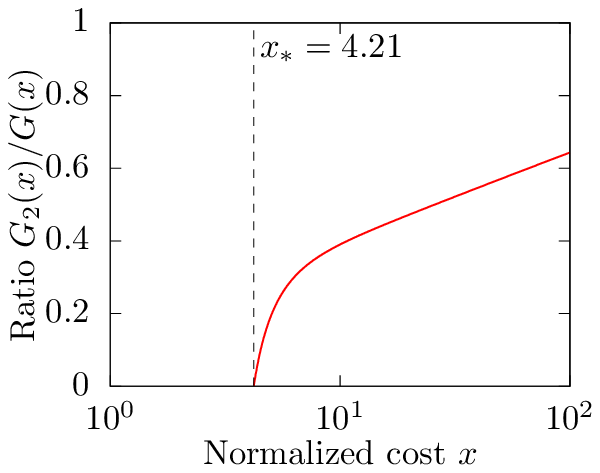}
\caption{
(a) Decomposition of $G(x)$ into the lognormal part $G_1(x)$ and power-law-like part $G_2(x)$.
The parameter values are the same as those in Fig.~\ref{fig4}: $(\mu,\sigma)=(-0.84, 1.24)$, $x_*=4.21$, and $\beta=5.28$.
(b) The ratio $G_2(x)/G(x)$ increases with increasing $x$, which suggests that the power-law-like part $G_2(x)$ is dominant in the tail.
}
\label{fig5}
\end{figure}

As outlined in Sec.~\ref{sec3} [see Eqs.~\eqref{eq8} and \eqref{eq10}],
$G(x)$ for $x\ge x_*$ can be decomposed into two parts $G(x)=G_1(x)+G_2(x)$,
where $G_1(x)$ is a lognormal function, and $G_2(x)$ is a power-law-like function (not an exact power law due to the $\erf$ factor).
Figure~\ref{fig5}(a) shows $G_1(x)$ and $G_2(x)$ for $(\mu, \sigma)=(-0.84, 1.24)$, $x_*=4.21$, and $\beta=5.28$ (same values as in Fig.~\ref{fig4}).
In this case, the magnitude of $G_2(x)$ is greater than that of $G_1(x)$ for $x>25.8$, and the power-law-like part $G_2$ is dominant in the tail.
Specifically, as Fig.~\ref{fig5}(b) shows, the ratio $G_2(x)/G(x)$ represents an increasing function.
Furthermore, using the asymptotic expansion $1-\erf(z)\simeq\exp(-z^2)/(z\sqrt{\pi})$, $G_2(x)/G(x)\to1$ can be presented as $x\to\infty$ for any parameter value.

\begin{figure}[t!]\centering
\raisebox{42mm}{(a)}\includegraphics[scale=0.9]{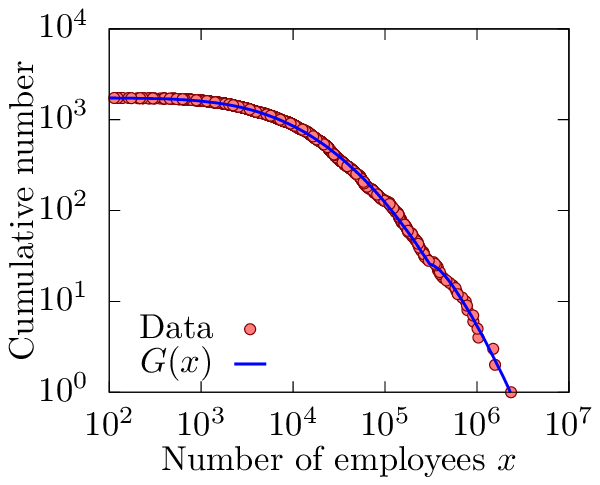}
\hspace{1cm}
\raisebox{42mm}{(b)}\includegraphics[scale=0.9]{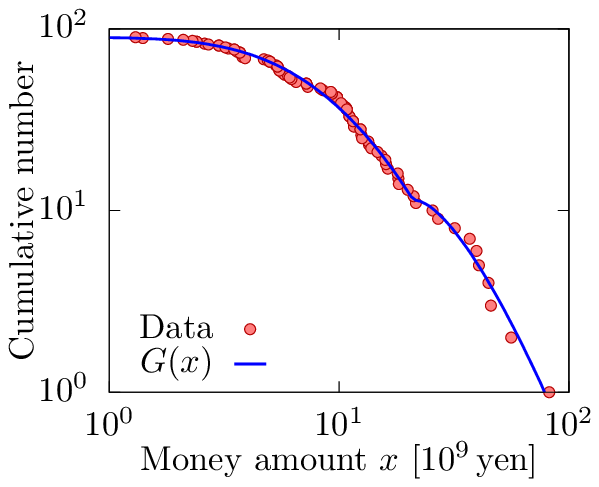}
\caption{
(a) Cumulative number of employees in municipalities ($N=1737$) in Japan in 2014.
The solid curve represents $G(x)$ for $(\mu,\sigma)=(9.17, 1.59)$, $x_*=3.09\times10^5$, and $\beta=4.20$.
(b) Cumulative number of management expenses grants for the national universities ($N=90$) in Japan in 2019.
The solid curve corresponds to $G(x)$ for $(\mu, \sigma)=(2.11, 0.82)$, $x_*=\SI{21e9}{yen}$, and $\beta=3.80$.
}
\label{fig6}
\end{figure}

In addition to the troubleshooting costs, we present the empirical distributions of other social data.
Figure~\ref{fig6}(a) shows the distribution of the number of employees aggregated according to a municipality ($N=1737$) in Japan~\cite{SSDE} in the form of the cumulative number.
Here, the solid curve represents $G(x)$ for $(\mu,\sigma)=(9.17, 1.59)$, $x_*=3.09\times10^5$, and $\beta=4.20$.
We believe that the lognormal behavior in the $x<x_*$ region is related to the lognormal population distribution~\cite{Sasaki}, and that the heavy tail likely reflects the extreme concentration of companies in metropolises, such as Tokyo and Osaka.
Meanwhile, Fig.~\ref{fig6}(b) shows the cumulative number (points) of the management expenses grants for 90 national universities in Japan in 2019~\cite{MEXT} and $G(x)$ for $(\mu, \sigma)=(2.11, 0.82)$, $x_*=\SI{21e9}{yen}$, and $\beta=3.80$ (solid curve).
Here, $12$ universities have $x$ greater than $x_*$.
While it is not clear whether these data are generated by the same statistical mechanism as that outlined in this study in terms of the troubleshooting costs,
we believe that $G(x)$ effectively expresses the empirical distribution; thus, providing some insight into these phenomena.

\begin{figure}[t!]\centering
\raisebox{38mm}{(a)}
\includegraphics[scale=0.85]{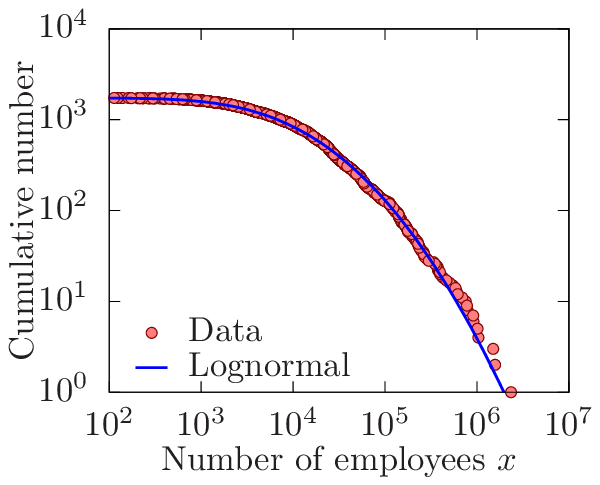}
\hspace{5mm}
\includegraphics[scale=0.85]{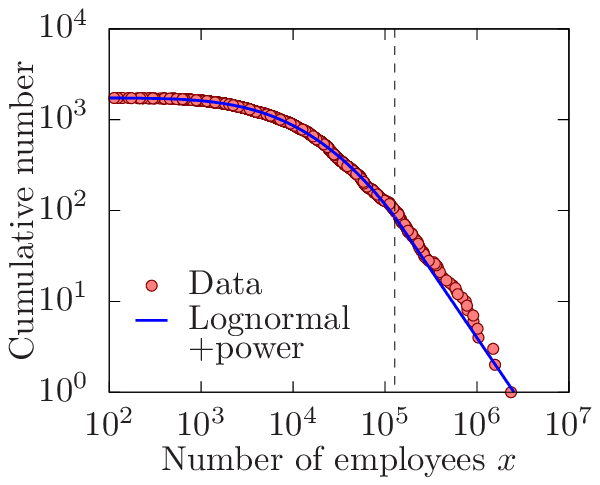}
\hspace{5mm}
\includegraphics[scale=0.85]{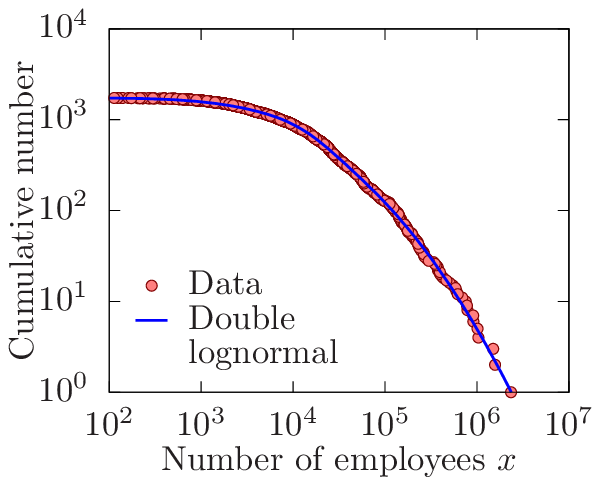}\\[5mm]
\raisebox{38mm}{(b)}
\includegraphics[scale=0.85]{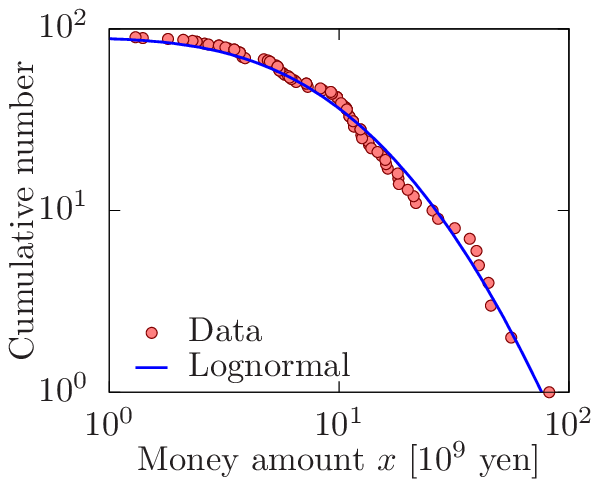}
\hspace{5mm}
\includegraphics[scale=0.85]{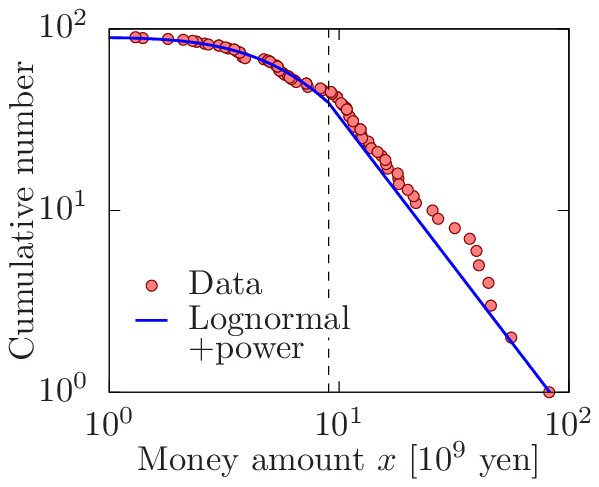}
\hspace{5mm}
\includegraphics[scale=0.85]{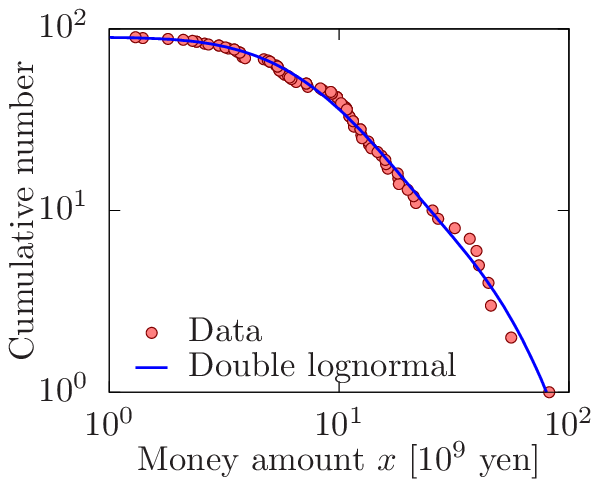}
\caption{
Results of the curve fitting for (a) the number of employees and (b) the management expenses grants for the Japanese national universities
by the lognormal distribution (left), the lognormal distribution and power-law tail (center), and the double lognormal distribution (right).
The vertical dashed line in the center graphs indicates the border between the lognormal and power-law distributions.
}
\label{fig7}
\end{figure}

Next, we compare the proposed distribution $G(x)$ with other distribution functions.
We selected the following three distributions that have been frequently used in the analysis of empirical data:
the lognormal distribution $F(x)$ [shown in Eq.~\eqref{eq1b}], 
the lognormal distribution with power-law tail, 
and the double lognormal distribution (a mixture of two lognormal distributions) given by
\[
\frac{\alpha}{2}\left[1-\erf\left(\frac{\ln x-\mu_1}{\sqrt{2}\sigma_1}\right)\right]
+\frac{1-\alpha}{2}\left[1-\erf\left(\frac{\ln x-\mu_2}{\sqrt{2}\sigma_2}\right)\right],
\]
where $\mu_1, \sigma_1, \mu_2, \sigma_2$, and $\alpha$ are the fitting parameters.
Figure~\ref{fig7} shows the cumulative number of employees in municipalities (a) and management expenses grants for the Japanese national universities (b) (the datasets are the same as with Fig.~\ref{fig6}), and these three distributions.

To obtain the center graphs of Fig.~\ref{fig7}, we performed the following procedure.
First, the power-law exponent and the lower bound of the power-law distribution were determined using the method outlined in Ref.~\cite{Clauset}, and we regarded this lower bound as the crossover point of the power-law and lognormal distributions.
Next, the lognormal fitting was performed in the range up to the crossover point.
Finally, the coefficient of the power-law distribution was computed so that the power-law and lognormal functions were continuously connected.

\color{black}


To evaluate the quality of curve fitting using the four distributions (the lognormal distribution, the lognormal distribution with power-law tail, the double lognormal distribution, and the proposed distribution), we measured the distance between the empirical distribution and each candidate distribution.
Among a wide variety of distances or metrics between probability distributions, we selected two simple ones:
the Kolmogorov-Smirnov statistic $D$ and the first Wasserstein distance $W_1$.
If the cumulative distributions of the empirical data and the fitting result are respectively written as $F_\mathrm{emp}$ and $F_\mathrm{fit}$, these two distances are defined as
\begin{align*}
D(F_\mathrm{fit}, F_\mathrm{emp})&=\sup_{0<x<\infty}|F_\mathrm{fit}(x)-F_\mathrm{emp}(x)|,\\
W_1(F_\mathrm{fit}, F_\mathrm{emp})&=\int_0^\infty |F_\mathrm{fit}(x)-F_\mathrm{emp}(x)|dx.
%
\end{align*}
These distances take non-negative values, and the small distances indicate that the estimated distribution is close to the empirical one.

\begin{table}[t!]\centering
\caption{
Distances between the empirical data and the fitting distributions.
The datasets are (a) number of employees, (b) management expenses grants, and (c) troubleshooting costs.
The relative values to the case of the proposed distribution $G$ are shown.
}
(a) Number of employees\\
\begin{tabular}{l|ccc}
\hline
 & \multirow{2}{*}{Lognormal} & Lognormal & Double\\
 & & +power & lognormal\\
\hline
Relative $D$ & 1.49 & 1.00 & 1.00\\
Relative $W_1$ & 1.78 & 1.78 & 0.95\\
\hline
\end{tabular}
\\[0.5\baselineskip]
(b) Management expenses grants\\
\begin{tabular}{l|ccc}
\hline
 & \multirow{2}{*}{Lognormal} & Lognormal & Double\\
 & & +power & lognormal\\
\hline
Relative $D$ & 1.21 & 1.65 & 1.12\\
Relative $W_1$ & 1.87 & 2.62 & 1.28\\
\hline
\end{tabular}
\\[0.5\baselineskip]
(c) Troubleshooting costs\\
\begin{tabular}{l|ccc}
\hline
 & \multirow{2}{*}{Lognormal} & Lognormal & Double\\
 & & +power & lognormal\\
\hline
Relative $D$ & 1.49 & 1.00 & 1.00\\
Relative $W_1$ & 1.78 & 1.78 & 0.95\\
\hline
\end{tabular}
\label{tbl1}
\end{table}

The calculation result of $W_1$ and $D$ for the lognormal distribution, lognormal distribution with power-law tail, and double lognormal distribution in Fig.~\ref{fig7}, in comparison with the proposed distribution $G$ in Figs.~\ref{fig6} and \ref{fig4}, is presented in Table~\ref{tbl1}.
The relative values (the distance between each distribution and the empirical distribution divided by the case of the proposed distribution $G$) are shown in the table; the value greater than unity represents that the distribution $G$ is closer to the empirical distribution than the corresponding distribution with respect to the corresponding distance.

\begin{figure}[t!]\centering
\includegraphics[scale=0.9]{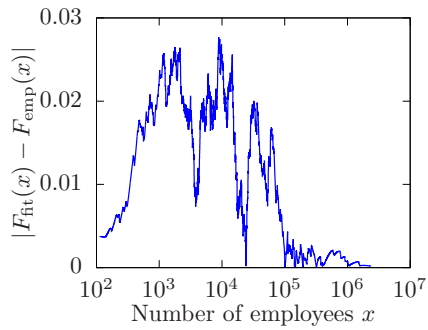}
\caption{
The absolute difference $|F_\mathrm{fit}(x)-F_\mathrm{emp}(x)|$ for the case of the distribution of the number of employees and the double lognormal distribution [corresponding to the right panel of Fig.~\ref{fig7}(a)].
The differences in the tail ($x\gtrsim10^5$) are small compared with those in the body ($x\lesssim10^5$).
}
\label{fig8}
\end{figure}

In the data analysis of complex systems including social ones, the tail of empirical distributions has often been focused on.
However, the above statistical distances are not suitable for the evaluation of the tail behavior.
In fact, Fig.~\ref{fig8} shows the absolute difference $|F_\mathrm{fit}(x)-F_\mathrm{emp}(x)|$ of the empirical distribution for the number of employees and its double-lognormal fitting result [Fig.~\ref{fig7}(a) right] as a function of $x$, indicating that large differences occur in the region $x\lesssim10^5$.
(A similar tendency is found in the other two datasets and other distributions.)
Hence, $W_1$ and $D$ are dominantly affected by the body part, and the contribution of differences in the tail is relatively small.
The four distributions (lognormal, lognormal with power-law tail, double lognormal, and $G$) have lognormal forms for small $x$, and similar $W_1$ and $D$ values in Table~\ref{tbl1} are considered to be attributed to this common lognormality.

To magnify the difference in the tail, we propose two distances:
\[
\Delta_1(F_\mathrm{fit}, F_\mathrm{emp})=\int_{x_\mathrm{min}}^{x_\mathrm{max}} |\ln F_\mathrm{fit}(x) - \ln F_\mathrm{emp}(x)| dx
\]
and
\[
\Delta_2(F_\mathrm{fit}, F_\mathrm{emp})=\int_{x_\mathrm{min}}^{x_\mathrm{max}} \frac{|F_\mathrm{fit}(x) - F_\mathrm{emp}(x)|}{F_\mathrm{emp}(x)} dx,
\]
where $x_\mathrm{min}$ and $x_\mathrm{max}$ are the minimum and maximum values of the empirical data, respectively.
(The upper limit of the integral is set to $x_\mathrm{max}$ in order to avoid the singularity $F_\mathrm{emp}(x)=0$ for $x>x_\mathrm{max}$.)
Like other statistical distances, $\Delta_1$ and $\Delta_2$ are always nonnegative and they become smaller as the estimated distribution $F_\mathrm{fit}$ is closer to the empirical one $F_\mathrm{emp}$.

\begin{table}[t!]\centering
\caption{
Distances $\Delta_1$ and $\Delta_2$ between the empirical data and the fitting distributions.
The datasets are (a) number of employees, (b) management expenses grants, and (c) troubleshooting costs.
The relative values to the case of the proposed distribution $G$ are shown.
}
(a) Number of employees\\
\begin{tabular}{l|ccc}
\hline
 & \multirow{2}{*}{Lognormal} & Lognormal & Double\\
 & & +power & lognormal\\
\hline
Relative $\Delta_1$ & 4.58 & 2.96 & 1.13\\
Relative $\Delta_2$ & 3.88 & 2.56 & 1.09\\
\hline
\end{tabular}
\\[0.5\baselineskip]
(b) Management expenses grants\\
\begin{tabular}{l|ccc}
\hline
 & \multirow{2}{*}{Lognormal} & Lognormal & Double\\
 & & +power & lognormal\\
\hline
Relative $\Delta_1$ & 1.42 & 2.38 & 1.26\\
Relative $\Delta_2$ & 1.35 & 1.95 & 1.27\\
\hline
\end{tabular}
\\[0.5\baselineskip]
(c) Troubleshooting costs\\
\begin{tabular}{l|ccc}
\hline
 & \multirow{2}{*}{Lognormal} & Lognormal & Double\\
 & & +power & lognormal\\
\hline
Relative $\Delta_1$ & 1.87 & 3.06 & 1.08\\
Relative $\Delta_2$ & 1.71 & 4.32 & 1.12\\
\hline
\end{tabular}
\label{tbl2}
\end{table}

The quality of curve fitting based on $\Delta_1$ and $\Delta_2$ is presented in Table~\ref{tbl2}.
As in Table~\ref{tbl1}, the relative values of the three distributions in Fig.~\ref{fig7} against the distribution $G$ in Figs.~\ref{fig6} and \ref{fig4} are shown in this table.
The proposed distribution $G$ describes the tail of empirical distributions better than (or at least comparable to) the other three distributions (lognormal, lognormal with power-law tail, and double lognormal).

In addition to the quantitative aspect, another advantage of $G$ is that it possesses a mechanism (stochastic process) different from the other three distributions.
\color{black}
That is, the agreement of the empirical dataset with the proposed distribution implies that the data was generated under the effect of the proposed mechanism, namely a multiplicative stochastic process with a sample-dependent lower bound.
This conclusion is based on only the three specific datasets shown in Figs.~\ref{fig4} and \ref{fig6}, and general discussion on the quality of each candidate distribution is the subject for a future study.

\section{Discussion}
The empirical data of troubleshooting costs outlined in this paper is normalized in terms of the average.
Here, we analyze the effect of this normalization.
If we set $m$ as the average cost, the raw value $\tilde{x}$ and the normalized value $x$ satisfy $x=\tilde{x}/m$.
The threshold $x_*$ is transformed into $\tilde{x_*}=mx_*$, since $x\lessgtr x_*$ changes to $\tilde{x}\lessgtr mx_*=\tilde{x_*}$,
while the parameter $\mu$ in $G(x)$ appears in the form of $\ln x-\mu$ or $\ln x_*-\mu$ at all times
[the factor $\exp(\mu\beta)x^{-\beta}$ is reduced to $\exp(-\beta(\ln x-\mu))$].
Therefore, the parameter $\mu$ becomes $\tilde{\mu}=\mu+\ln m$.
It should be noted that the parameters $\sigma$ and $\beta$ in $G$ do not change due to the normalization.
In conclusion, writing the parameters explicitly as $G(x; \mu, \sigma, x_*, \beta)$,
we have $G(x; \mu, \sigma, x_*, \beta)=G(\tilde{x}; \mu+\ln m, \sigma, mx_*, \beta)$.
The same dependence of the parameters holds even if the constant $m$ is not equal to the average value.

The distributions $g(x)$ and $G(x)$ are derived based on the assumption that the lower bound of the multiplicative process is lognormally distributed.
We can generalize these results by replacing the lognormal weight $f(a)$ described in Eq.~\eqref{eq6} with alternative probability distributions.
Here, when a probability density $h$ and its cumulative distribution $H$ are used instead of the lognormal distribution $f$ and $F$,
we obtain $G_1(x)=H(x)$ and
\[
G_2(x)=x^{-\beta}\int_{x_*}^x a^\beta h(a) da,
\]
which are similar to Eqs.~\eqref{eq8} and \eqref{eq9}.
The asymptotic power law $G(x)\simeq G_2(x)\sim x^{-\beta}$ generally holds if $h$ decays faster than $x^{-\beta-1}$.

The tail of an empirical probability distribution usually contains only a small amount of data.
For example, while the total number of cases is $N=2571$ in Fig.~\ref{fig4}, only $85$ cases (\SI{3.3}{\percent} of the total) lie in $x\ge x_*$.
Due to statistical fluctuations, the distribution tail is difficult to specify.
From Figs.~\ref{fig4} and \ref{fig6}, the proposed distribution $G$ fits well even up to the tail, which appears to increase reliability for the estimation of the asymptotic power-law exponent $\beta$.
By its construction, the proposed model, a multiplicative process with a sample-dependent lower bound, will be useful for a lognormal distribution with a fat tail, especially for a system whose dynamics or evolution rules change at a certain threshold value.
In fact, the troubleshooting cost, the number of employees, and the university management expenses grants analyzed in this study are real examples which follow the proposed distribution $G$, and this result suggests that the proposed mechanism may be embedded in the generation process of them.
Moreover, we expect that there exist other systems that can be well described by the proposed model.

We think that the proposed distribution $g$ or $G$ is an effective distribution but not the definitive one.
In general, the distribution function of a given dataset should be determined by examining several candidate distributions, and we hope that the distribution $G$ outlined in this paper becomes one of the possible choices.

\section*{Acknowledgments}
The authors are very grateful to Mr.\ Hirokazu Yanagawa and Mr.\ Takashi Bando for constructive discussion and permission to use data.
KY was supported by a Grant-in-Aid for Scientific Research (C) 19K03656 from Japan Society for the Promotion of Science.


\begin{thebibliography}{10}
\bibitem{Company} At the company's request, its name cannot be revealed, and the use of the absolute cost values was prohibited.
\end{thebibliography}
\end{document}